# New Symmetric and Planar Designs of Reversible Full-Adders/Subtractors in Quantum-Dot Cellular Automata

Moein Sarvaghad-Moghaddam[1], Ali A. Orouji[1,*]

[1] Department of Electrical and Computer Engineering, Semnan University, Semnan, Iran
Email: moeinsarvaghad@yahoo.com; aliaorouji@semnan.ac.ir

*Abstract*— **Quantum-dot Cellular Automata (QCA) is one of the emerging nanotechnologies, promising alternative to CMOS technology due to faster speed, smaller size, lower power consumption, higher scale integration and higher switching frequency. Also, power dissipation is the main limitation of all the nano electronics design techniques including the QCA. Researchers have proposed the various mechanisms to limit this problem. Among them, reversible computing is considered as the reliable solution to lower the power dissipation. On the other hand, adders are fundamental circuits for most digital systems. In this paper, Innovation is divided to three sections. In the first section, a method for converting irreversible functions to a reversible one is presented. This method has advantages such as: converting of irreversible functions to reversible one directly and as optimal. So, in this method, sub-optimal methods of using of conventional reversible blocks such as Toffoli and Fredkin are not used, having of minimum number of garbage outputs and so on. Then, Using the method, two new symmetric and planar designs of reversible full-adders are presented. In the second section, a new symmetric, planar and fault tolerant five-input majority gate is proposed. Based on the designed gate, a reversible full-adder are presented. Also, for this gate, a fault-tolerant analysis is proposed. And in the third section, three new 8-bit reversible full-adder/subtractors are designed based on full-adders/subtractors proposed in the second section. The results are indicative of the outperformance of the proposed designs in comparison to the best available ones in terms of area, complexity, delay, reversible/irreversible layout, and also in logic level in terms of garbage outputs, control inputs, number of majority and NOT gates.**

## I. INTRODUCTION

Energy dissipation is one of the major issues in present-day technology. Landauer [1] has proved that irreversible computation will have $k_B T \ln 2$ joules of dissipated energy due to loss of single bit information. Here, $k_B$ and $T$ holds for Boltzmann's constant and the computing temperature. In 1973, Bennett, showed that in order to avoid energy dissipation in a circuit it must be built from reversible circuits [2]. Reversible computing considers the relation between information dissipation and energy dissipation at the logical level. If a one-to-one mapping is established between the input and output vectors, reversible computation would be achieved at the logical level [2]. The one-to-one mapping is called bijective property. In addition, Fan-out is not possible in reversible systems. The unused outputs are used to maintain the reversibility of the circuits and are known as the garbage outputs. The constant inputs in the reversible circuits are called the ancilla.

Quantum-dot Cellular Automata (QCA) as a well-known technology is able to replace devices based on FET (Field Effect Transistor) on nano-scale [3-5]. Promising size density of several orders of magnitude smaller than CMOS, fast switching time and extremely low power, has caused QCA to become a very interesting topic of researches [6-8]. A QCA cell consists of four dots and two identical electrons. Each dot can be occupied by one of the two hopping electrons in a way that the electrons stay diagonally opposite owing to Columbic interaction. In QCA technology binary information is encoded by formations of electrons rather than current employed in CMOS. Two fundamental building blocks for QCA are the inverter and the majority gate [3]. In QCA Logic, clock is responsible for synchronization and control of information flow and also it provides power to run the circuit [9, 10].

However, it should be pointed out that the majority voting function is logically irreversible because the information on the minority input is lost during the computation. It has been shown in the literature that in QCA different clocking arrangements can be used for reversible computing. By using the clocking scheme is referred to as Bennett clocking could offer a practical realization of reversible computing using QCA. It has been shown by direct calculation that with Bennett clocking, energy dissipation per switching event is much less than $kT\ln2$ for QCA circuits with devices such as MV and fan-out [11, 12].

Adders are fundamental circuits for most digital systems and several adder designs in QCA have been proposed [13-16] and a performance comparison was presented [13]. In some of full adder designs, 3-input majority gates were employed [17, 18] however recently 5-input ones have been used in order to reduce the number of cells as well as occupied area in designs [19-21]. In design proposed in [22], there are three three-input majority gates, two inverters and four clocking phases. In [23] a QCA full-adder is implemented only using three gates, two majority gates and one inverter. In this design has been used a five-input majority gate that it has a simpler design scheme in comparison to other previous design [22]. Five- input majority proposed in [23] has a cubic structure and maybe it is not simply feasible to fabricate. In [18] a new design for QCA full-adders is presented which revised the previous full-adder scheme. This design has only three clock phases.

C. S. Lent et al in 2005, proposed circuit design based on QCA in reversible logic [24]; then several authors proposed several reversible gates in QCA [25-28]. In [29], designing of a one-bit full adder has been investigated using a QCA implementation of Toffoli and Fredkin gates. Then, a full adder design with reversible QCA1 gates has been proposed. Also, in [30], a novel $3 \times 3$ reversible gate that is universal and testable has been presented. By using this gate, a new 8-bit full-adder is designed. Work in this paper is even worse into work [29] in terms of complexity and area and delay.

In this paper, innovation is divided to three section. In the first section, a method for converting irreversible functions to a reversible one is presented. This method has advantages such as: converting of irreversible functions to reversible one directly and as optimal. So, in this method, sub-optimal methods of using of conventional reversible blocks such as Toffoli and Fredkin are not used, having of minimum number of garbage outputs and garbage outputs do not create any additional gates (Majority and NOT gates). Then, Using the method, two new symmetric and planar designs of reversible full-adders are presented. In the second section, a new symmetric, planar and fault tolerant five-input majority gate is proposed. Based on the designed gate, a reversible full-adder are presented. Also, for this gate, a fault tolerant analysis is proposed [22]. And in the third section, three new 8-bit reversible full-adder/subtractors is designed based on full-adders/subtractors proposed in the second section. In comparison to the best existing implementations, the proposed designs have demonstrated significant improvements in terms of area, complexity, delay, reversible/irreversible layout, and also in logic level in terms of garbage outputs, control inputs, number of majority and NOT gates.

The rest of the paper is organized as follows: Section II present some related background materials. Section III introduces the proposed method in detail. Section IV show simulation results and comparison. Section V concludes the paper.

## II. BACKGROUND MATERIAL

In this section, basic concepts in QCA technology such as Quantum-dot cellular automata and QCA devices are explained.

### A. Quantum-Dot Cellular Automata

As shown in Fig. 1.a, a standard QCA cell is composed of four quantum-dots and two excess electrons located at the corners of a square. According to the existing Coulombic interaction between the electronic charges, they can occupy diagonal antipodal sites through tunneling junctions. Therefore, a single QCA cell can accept two completely polarized states called cell polarization $P = +1$ (binary '1' state) and $P = -1$ (binary '0' state) as shown in Fig. 1.b.

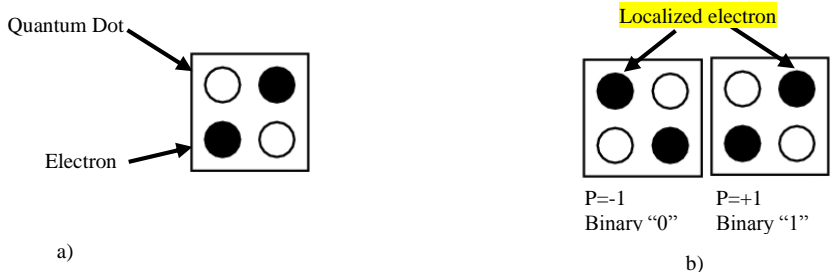

Fig. 1. a) Structure of a QCA cell with four quantum dots. b) QCA cell with two different polarizations.

### B. QCA Logic Devices

The fundamental QCA logic devices include a QCA wire, QCA inverter, and QCA majority gate. A 90° QCA wire is just a line of QCA cells. The wire is driven at the input cell by a cell with a fixed/held polarization [31]. The signal propagates along the wire from left to right when excited from the leftmost cell. A QCA wire can also be built with cells rotated by 45°. This kind of wire propagates the input signal in odd cells and inversion of the input signal in even cells [32, 33]. A QCA majority gate can perform a three-input logic function as given in (1), where $A$, $B$, and $C$ are the three inputs.

$$M(A, B, C) = AB + BC + CA. \tag{1}$$

By forcing one of the three inputs of the majority gate to a constant logic "0" or a "1" the majority gate can be used to perform AND/OR operations as shown in the following equations:

$$M(A, B, 0) = AB, \qquad M(A, B, 1) = A + B. \tag{2}$$

Fig. 2 demonstrates a QCA wire in 90° and 45°, inverter gate, and majority gate, respectively.

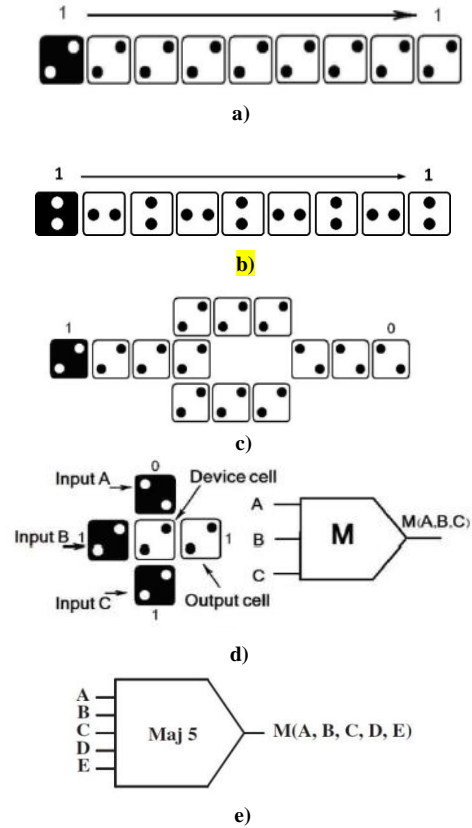

Fig. 2. Representation of a) 90° QCA wire b) 45° QCA wire c) QCA inverter gate d) QCA majority gate e) QCA five-input majority gate.

In QCA, there are two crossover options. They are coplanar crossings and multilayer crossovers. In the first way, one quantum wire with a 45-degree turn passes over a regular quantum wire without any interference. This approach is shown in Fig. 3.a). In the other approach, multi-layered structures are used for the passage of the quantum wires over each other. This structure is shown in Fig. 3.b).

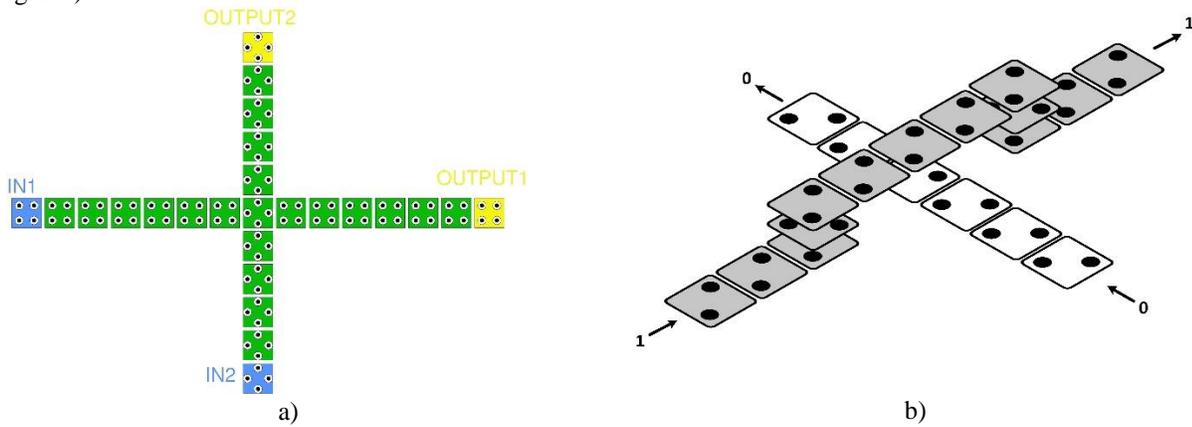

Fig. 3: a) coplanar crossing wire b) multilayer crossing wire

## C. Five-Input Majority Gate

A five pins majority gate must have five inputs and one output. Fig. 2.e illustrates a schematic of five-input majority gate. The majority voting logic function can be expressed in terms of fundamental Boolean operator as shown in (3).

$$M(A, B, C, D, E) = ABC + ABD + ABE + ACD + ACE + ADE + BCD + BCE + BDE + CDE \qquad (3)$$

A three-input AND gate and also a three-input OR gate can be implemented using this majority gate. These functions are as (4).

$$M(A, B, C, 0,0) = ABC,$$ 
(4)
$$M(A, B, C, 1,1) = A + B + C.$$

### D. QCA Clocking

In VLSI circuits, timing is controlled by a reference signal which is called clock but in QCA, the information storage and erasure in the cell are controlled by the clock. Clocking is required for QCA circuit to synchronize and information flow control. Two strategies have been considered for clocking QCA [11]: 1) Landauer clocking: has a logically irreversible "erase" operation 2) Bennett clocking: has a logically reversible "copy-then-erase" operation [34]. It has been shown that erasure without copying requires an amount of dissipated energy in the order of the signal energy. This scheme could offer a practical realization of reversible computing using QCA.

Generally, four multi-phase clocking signals applied as shown in Fig. 4.a is called Landauer type [35]. During a complete cycle, each zone goes through the four phases. This zones must follow a particular order namely $C_0 \rightarrow C_1 \rightarrow C_2 \rightarrow C_3$. Four phases are called Switch, Hold, Release, and Relax [10]. In switch phase, QCA cell is starting to move from un-polarized state to polarize state and the barriers of the dots is raised. The electrons are started tunneling through dots as the dots are influenced by the electron of its neighbor cell. In Hold phase, barrier of the cell is in the high value, electron can't tunnel through dots and cell maintains their current states i.e. fixed polarization. In release phase, barrier is lowered, electron can tunnel through dots and states of the cell become unpolarized. In relax phase barrier remains lowered and cell stay in un-polarized state. Another type of clocking signals for reversible circuit is Bennett-type clocking [34]. The waveform of Bennett clocking is shown in Fig. 4.b. The principle of this clocking is to first compute the results by latching the cell array from input to output and then uncompute by latching array to relax to an unpolarized state from output to input.

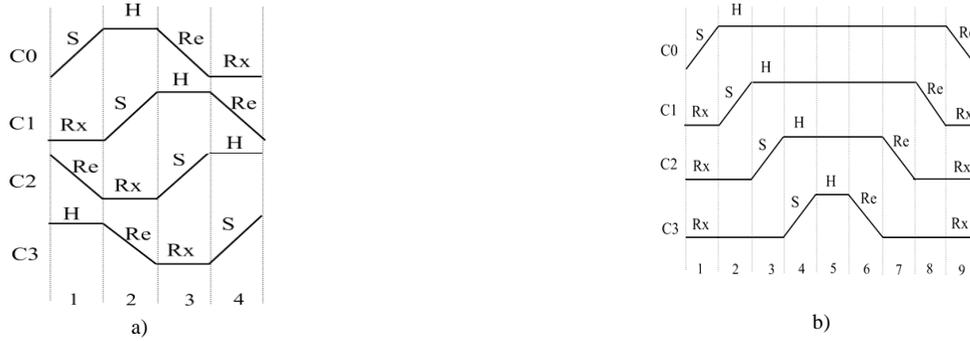

Fig. 4. a) Laudauer clocking waveform. b) Bennett clocking waveform. S: switch phase, H: hold phase, Re: release phase, Rx: relax phase.

## III. PROPOSED METHOD

In this section, first a method is introduced for converting multi–output irreversible functions to reversible ones. Then, according to it, new designs of reversible full adder are proposed. In addition, a new structure of five-input majority gate is presented that based on it, other design of reversible full adder is presented.

### A. Method for converting multi-output functions to reversible functions

As stated in pervious sections, a reversible function has the following features:
1- There is a one-to-one mapping between inputs and outputs.
2- Feedback is not allowed.
3- Fan-out is not allowed.

But, as stated in [11, 12] features 2 and 3 is not necessary for creating reversible functions in QCA technology. So, for converting the multi-output functions to the reversible functions, only, one to one mapping between inputs and outputs must be created. Also, it is not necessary that the number of inputs and outputs be same. To this end, first, common states in outputs are marked, then, input columns are compared with outputs and input column that can remove the most common states in outputs is selected and added as a new column in output. This work is repeated until common states in outputs is deleted fully. By using this method, directly, can create each type reversible functions with minimum complexity, area, garbage outputs, control inputs and delay while the another methods such as [29, 30, 36] have used common reversible blocks such as Toffoli and Fredkin gates or building of reversible blocks for designing another reversible functions that this methods are not optimal. Our method has the minimum number of garbage outputs. As in this method, garbage outputs are added one to one for removing the most common states. Another feature of this method is that garbage outputs do not create any additional gates (Majority and NOT gates) for converting irreversible functions to reversible one. So, the number of majority gates are optimal and are as same with irreversible state. So, it has minimum complexity, delay and area. Control inputs are redundancy inputs used for create AND/OR gates. For a reversible function, these inputs must be minimum for reducing power consumption. In the next sections, we design new reversible functions using this method in QCA technology.

*1) The proposed reversible full-adder*

By using proposed method for creating reversible functions, a new design of reversible full-adder is proposed. Specification function of this full-adder is shown in Table I. as shown in this table, the first, common states in outputs are determined (showed with gray color in Table I). Two columns $a$ and $b$ are added as output columns or garbage outputs. These two garbage outputs $Gar1$ and $Gar2$ are selected according to the most removing of common states in specification function and create a one to one mapping between inputs and outputs. In this table, there are three common states 01 and three common states 10. By adding $Gar1$ output that is as same with $a$ column in input ($Gar1 = a$), one common state in 01 state and one common state in 10 state are removed. Then, by adding Gar2 output as same with $c_{in}$ column in input ($Gar2 = c_{in}$), all common states in output are removed and one to one mapping between input and output are created.

TABLE I
DESIGNING OF NEW REVERSIBLE FULL-ADDER

| a | b | $c_{in}$ | Cout | Sum | Gar1 | Gar2 |
|---|---|---|---|---|---|---|
| 0 | 0 | 0 | 0 | 0 | 0 | 0 |
| 0 | 0 | 1 | 0 | 1 | 0 | 1 |
| 0 | 1 | 0 | 0 | 1 | 0 | 0 |
| 0 | 1 | 1 | 1 | 0 | 0 | 1 |
| 1 | 0 | 0 | 0 | 1 | 1 | 0 |
| 1 | 0 | 1 | 1 | 0 | 1 | 1 |
| 1 | 1 | 0 | 1 | 0 | 1 | 0 |
| 1 | 1 | 1 | 1 | 1 | 1 | 1 |

In the next step, for synthesizing each of outputs, method in [37] is used. As stated in above, garbage outputs are as same with inputs as the following:

$Gar1 = a,$

$Gar2 = c_{in}.$ (5)

$Cout$ and $Sum$ outputs can be synthesized as the following optimal forms with majority and NOT gates:

$Cout= ab+bc_{in}+ac_{in}= \underline{M(a,b,c_{in})},$ (6)

$Sum=abc_{in}+ab'c_{in}'+a'bc_{in}'+a'b'c_{in}=M(\underline{M(a,b,c_{in})},a, M(a',b,c_{in})).$

In Equation (6), common parts is underlined that as fan-out are used.

*Proof*   The proof of synthesis of $Sum$ output also can be done as follows (similar to [38]):

Using the majority function of $Cout$:

$Cout= a'b'+b'c_{in}'+a'c_{in}'= M(a',b',c_{in}').$

Then, $Sum$ output can be rewritten as:

$Sum=abc_{in}+ab'c_{in}'+a'bc_{in}'+a'b'c_{in}= a[(bc_{in}+a'c_{in}+a'b)+ (b'c_{in}'+a'b'+a'c_{in}')] + a'bc_{in}' + a'b'c_{in}= a[(bc_{in}+a'c_{in}+a'b) + (b'c_{in}'+a'b'+a'c_{in}')] + (a'b+a'c)(a'c_{in}'+a'b') = a[(bc_{in}+a'c_{in}+a'b) + (b'c_{in}'+a'b'+a'c_{in}')] + (a'b+a'c_{in}+bc_{in})(a'c_{in}'+a'b'+b'c_{in}') = a(M(a',b,c_{in})+M(a',b',c_{in}')) + M(a',b,c_{in})M(a',b',c_{in}') = M(M(a,b,c_{in})', a,M(a',b,c_{in})).$

Also Equation (6) can be rewrite as two other equations as the following:

$Sum = M(\underline{Cout}',b,M(a,b',c_{in})),$

$Sum= M(\underline{Cout},c_{in},M(a,b,c_{in}')).$ (7)

Three another form for synthesizing Sum output can be proposed as the following:

$Sum= M(M(a,b,c_{in}'),M(a,b',c_{in}),a'),$ (8)

$Sum= M(M(a,b,c_{in}'),M(a',b',c_{in}),b'),$

$Sum= M(M(a',b,c_{in}),M(a,b',c_{in}),c'.$

*Proof*   The proof of synthesizing of $Sum$ output in the Eq. 8 is as follows:

$Sum=ab\ c_{in} +ab'\ c_{in}'+a'b\ c_{in}'+a'b'\ c_{in} = a'(b\ c_{in}' + b'\ c_{in}) + (ab\ c_{in} + ab'\ c_{in}') = a'[(b\ c_{in}' + ab + a\ c_{in}) + (b'\ c_{in} + ab' + a\ c_{in})] + (ab + a\ c_{in}')(ab' + a\ c_{in}) = a'[(b\ c_{in}' + ab + a\ c_{in}) + (b'\ c_{in} + ab' + a\ c_{in})] + (ab + a\ c_{in}' + b\ c_{in}')(ab' + a\ c_{in} + b'\ c_{in}) = a'(M(a,b,\ c_{in}') + M(a,b',\ c_{in})) + M(a,b,\ c_{in}')M(a,b',\ c_{in}) = M(M(a,b,\ c_{in}'),\ M(a,b',\ c_{in}),\ a')$

Fig. 5 shows the schematic of the proposed symmetric irreversible full adder according to Eq. 6 without garbage outputs. As shown in this figure, this topology is symmetric and we use from this new schematic for improving QCA layout.

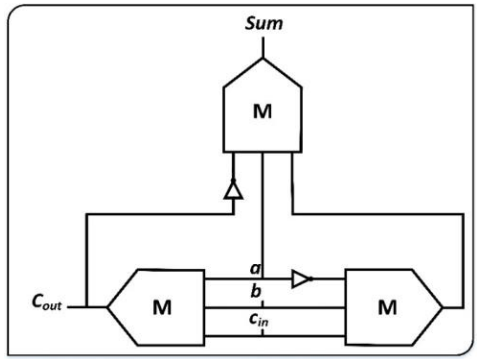

Fig. 5. Schematic of irreversible symmetric full-adder

According to Fig. 5 and Table I, the new layout of the proposed symmetric reversible full-adder is presented in Fig. 6. Due to symmetric structure, in designing of this layout, three clock phase are used that this feature is caused circuit delay is reduced. Also, this layout is planar and in one layer is implemented.

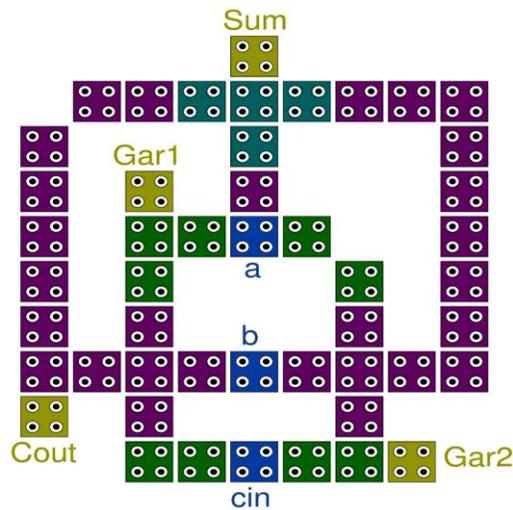

Fig. 6. Illustration of layout of proposed symmetric full-adder.

### 2) The proposed reversible full-adder/subtractor

In this section, a new design of reversible full-adder/subtractor is presented according to method proposed for converting irreversible functions to reversible one. Designing of specification function is shown in Table II. In this table, outputs of $Cout$, $Sum/Sub$ and $Bout$ are considered as the main outputs and $Gar1$ is garbage output that is added for reversibility. In fact, in this specification function instead of using of two redundant Outputs (garbage output) for converting function to reversibility, one of garbage outputs in Table I, is used as the main output for doing subtraction. In the other words, we used garbage output for specific aims and increase performance of reversible function. Also, in this design, outputs $Sum$ and $Sub$ is considered as one output and outputs $Cout$ and $Bout$ is considered as separate. So, this feature is caused that in this structure, operations of Adding and Subtracting is done as simultaneously. As a result, in this stage, also, propagation delay is reduced. Then, common states in output are specified (showed with gray color in Table II). Then as comparing with input columns, a column with the most removing of common states in output is selected. In this function, garbage output is selected as same with input column $a$. With adding this column and reviewing of output states is determined that one to one mapping between input and output have been created. The next step is synthesizing of outputs. $Cout$ and $Sum/sub$ output are synthesized as Eq. 6. Outputs of $Bout$ and Gar1 can obtain as the following:

$Bout=\underline{M(a',b,c)}.$                                               (9)

$Gar=a.$

where

$Bout$ output is common with $Sum$ in term $M(a',b,c)$. Schematic of this block is shown in Fig. 7.a).

According to Table II and Fig. 5, layout of symmetric reversible full-adder/subtractor is presented in Fig. 7.b). As shown in this figure and Equations (6) and (9), in this structure, we use garbage outputs that are redundant outputs as the main output for specific application (subtractor). Then for reducing the number of majority gates, we use $Sub$ output as a fanout in $Sum$ output. So, this output does not add any additional gate and structure is optimal. In addition, in this layout, symmetric introduced in Fig. 5 exist and this feature is caused that layout be optimal in terms of occupational area and cell count. Also, this layout is implemented in one layer similarly with layout shown in Fig. 6.

TABLE II
DESIGNING OF NEW REVERSIBLE FULL-ADDER/SUBTRACTOR

| $a$ | $b$ | $c_{in}$ | $Cout$ | $Sum/Sub$ | $Bout$ | $Gar1$ |
|---|---|---|---|---|---|---|
| 0 | 0 | 0 | 0 | 0 | 0 | 0 |
| 0 | 0 | 1 | 0 | 1 | 1 | 1 |
| 0 | 1 | 0 | 0 | 1 | 1 | 0 |
| 0 | 1 | 1 | 1 | 0 | 1 | 1 |
| 1 | 0 | 0 | 0 | 1 | 0 | 0 |
| 1 | 0 | 1 | 1 | 0 | 0 | 1 |
| 1 | 1 | 0 | 1 | 0 | 0 | 0 |
| 1 | 1 | 1 | 1 | 1 | 1 | 1 |

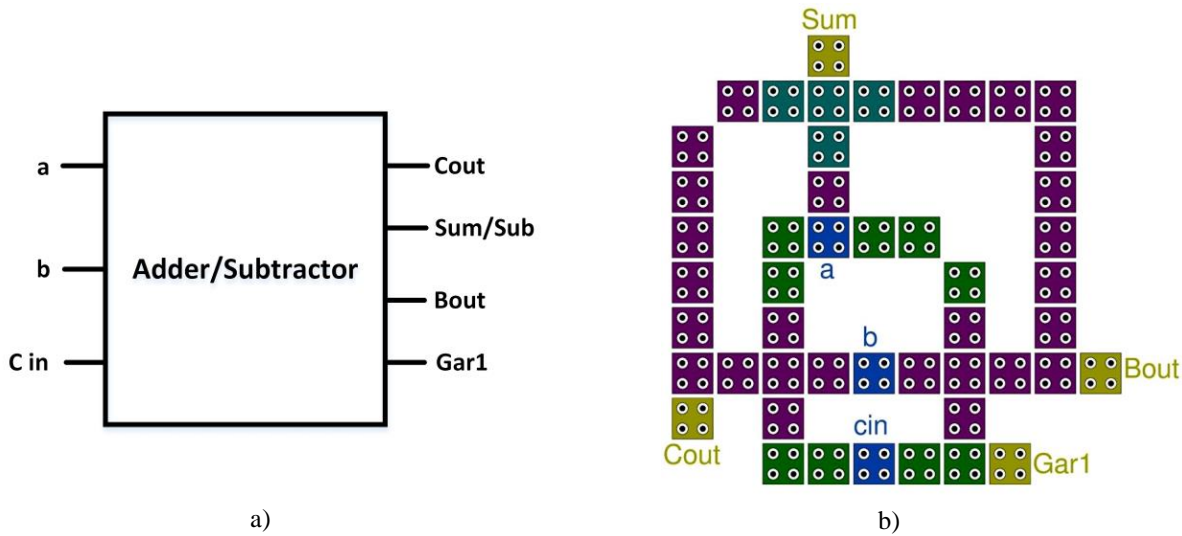

Fig. 7. Illustration of a) Schematic and b) Layout of proposed symmetric full-adder/subtractor

### B. A proposed symmetric five-input majority gate

In this section, a novel design of symmetric and planar structure of 5-input majority gate is presented as shown in Fig. 8. The gate is fault tolerant owing to its symmetric structure. Additionally, all input cells have an inverting effect on polarizations of device cells. The new five-input majority gate is composed of 10 cells, five input cells, one output cell and four device cells.

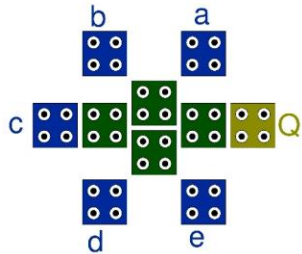

Fig. 8: Proposed symmetric five-input majority gate

A three-input AND gate and also a three-input OR gate can be implemented using this majority gate by fixing two of the five input cells to +1 or −1, respectively.

### 1) The proposed reversible full-adder using five-input majority gate

In this section, an efficient reversible QCA full adder is presented and implemented based on the proposed five-input majority gate. The schematic design and layout of the proposed irreversible full-adder is presented in Fig. 9. In addition, Fig. 10 illustrates the layout of the proposed reversible full-adder, which uses a planner five-input majority gate. In this figure, the layout is designed using Table I and schematic introduced in Fig. 9.a. Also, due to symmetric structure, three clock phase are used in this schematic. So, delay is reduced.

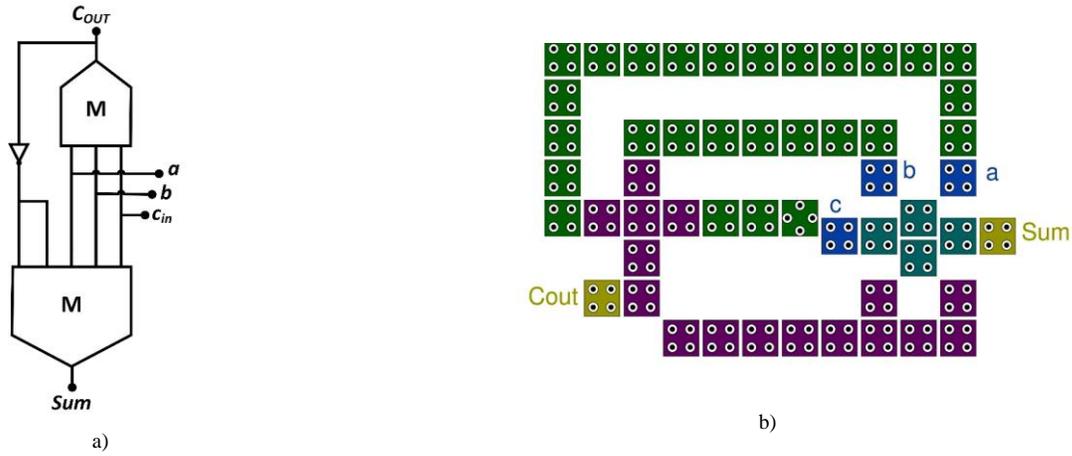

a)

b)

Fig. 9. Proposed irreversible QCA full-adder a) Schematic b) Layout.

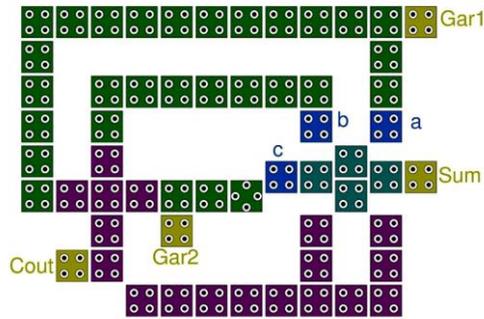

Fig. 10. Illustration of the layout of proposed reversible full-adder using five-input majority gate.

In practical, for applying inputs from outside of the device to inside of it, a method is using of a multilayer crossing wire technique as shown in Fig. 3 b). So, the proposed layouts can easily be implemented in practical.

### 2) Fault tolerance analysis

A fault-tolerant QCA gate should continue functioning correctly in cases where one or more of the cells are misaligned or misplaced. The first fault is due to displacement of cell from their intended location. The QCA cell displaced will be outside the radius of effect of its neighbor, so that no longer contributing to the interaction among the cells exists.

A typical maximum distance at which interaction exists is 40 to 60nm. To assess displacement and misalignment tolerance [19, 21], a 3-input AND gate based on our proposed gate is designed and cells $A$, $B$ and $C$ are displaced and misaligned in it. Fig. 11 shows the possible defects which may occur for cell $A$ (The green cells depict the possible locations of a displaced or misaligned $A$.). The obtained results are summarized in Table III.

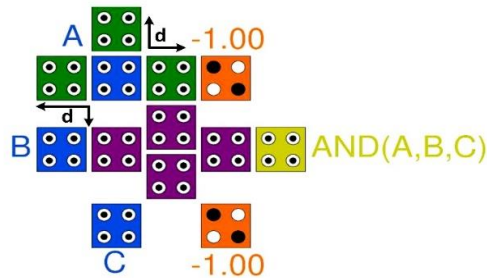

Fig. 11: Displacement and misalignment defects in the proposed three-input AND gate.

| CELL $C$ BEHAVES MIRRORLY SIMILAR TO CELL $A$ IN ALL DIRECTIONS | | | |
|---|---|---|---|
| MOVE $A$ TO NORTH | MOVE $A$ TO SOUTH | MOVE $A$ TO WEST | MOVE $A$ TO EAST |
| $0 \leq d \leq 5$ | $0 \leq d \leq 8$ | $0 \leq d \leq 6$ | $0 \leq d \leq 10$ |
| NORMAL FUNCTION | NORMAL FUNCTION | NORMAL FUNCTION | NORMAL FUNCTION |
| CELL $B$ | | | |
| MOVE $B$ TO NORTH | MOVE $B$ TO SOUTH | MOVE $B$ TO WEST | MOVE $B$ TO EAST |
| $0 \leq d \leq 6$ | $0 \leq d \leq 5$ | $0 \leq d \leq 5$ | NOT POSSIBLE |
| NORMAL FUNCTION | NORMAL FUNCTION | NORMAL FUNCTION | NORMAL FUNCTION |

## C. Reversible 8-bit full-adder/subtractor

In this section, three the 8-bit full-adder/subtractor are designed using adder/subtractor blocks proposed in pervious section. These circuits are a combinational circuit which performs arithmetic addition and subtraction with binary digits. This circuit is composed eight reversible full-adder/subtractor blocks as series connections. Fig. 12.a) and b) show schematics of reversible 8-bit full-adder and reversible 8-bit full-adder/subtractor using adder blocks proposed in pervious section, respectively. For creating a reversible 8-bit full-adder/subtractor in Fig. 12.b), multiplexer blocks $2 \times 1$ are used for detecting the addition or subtraction operations in each time. With selecting state zero, add operation is done and if state one is selected, subtraction operation is done.

The three QCA layout of 8-bit reversible full-adder/subtractor are shown in Fig. 13 a) and b) show layout of 8-bit reversible-full adders using adder blocks proposed in Fig.6 and Fig. 10, respectively. Fig. 13 c) show layout of 8-bit reversible full-adder/subtractor using adder/subtractor block proposed in Fig. 7. The proposed 8-bit reversible full-adders/subtractors consists of 570, 725,1040 cells covering an area of 0.55,0.68,1.12 $\mu m^2$ in Fig. 13 a) ,b) and c) respectively.

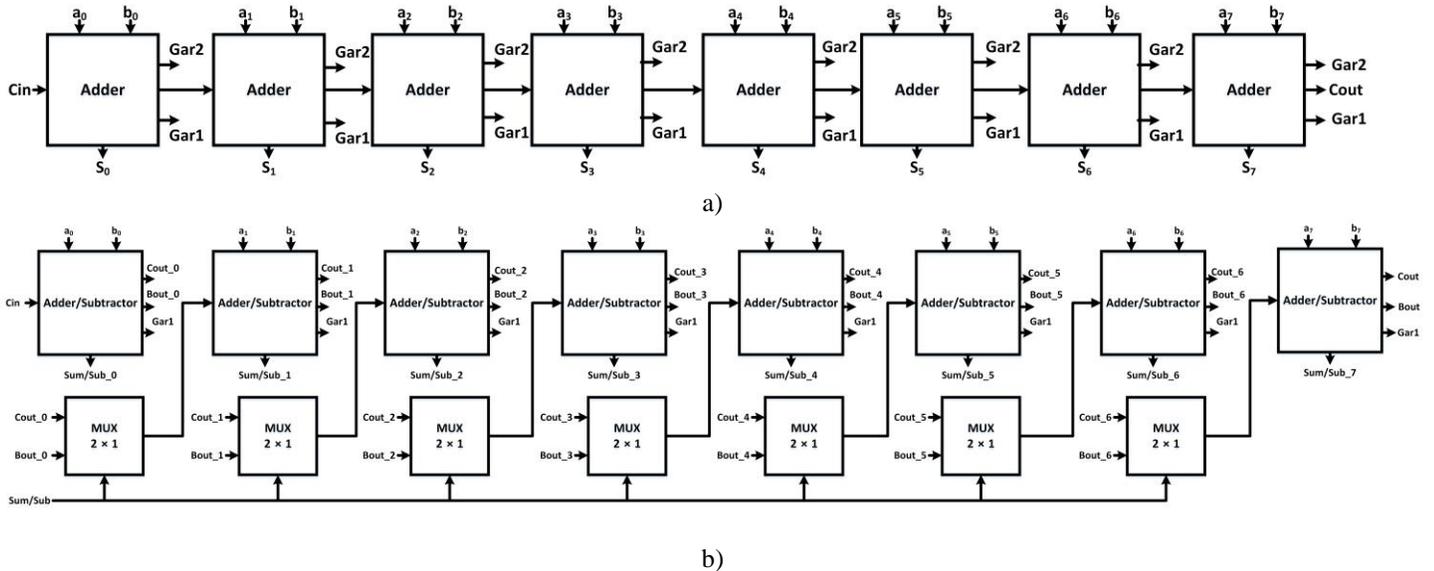

Fig. 3: a) Logic block of 8-bit reversible full-adder. b) Logic block of 8-bit reversible full-adder/subtractor

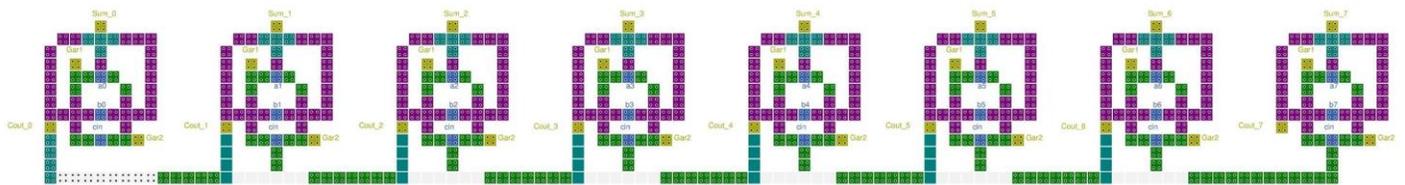

a)

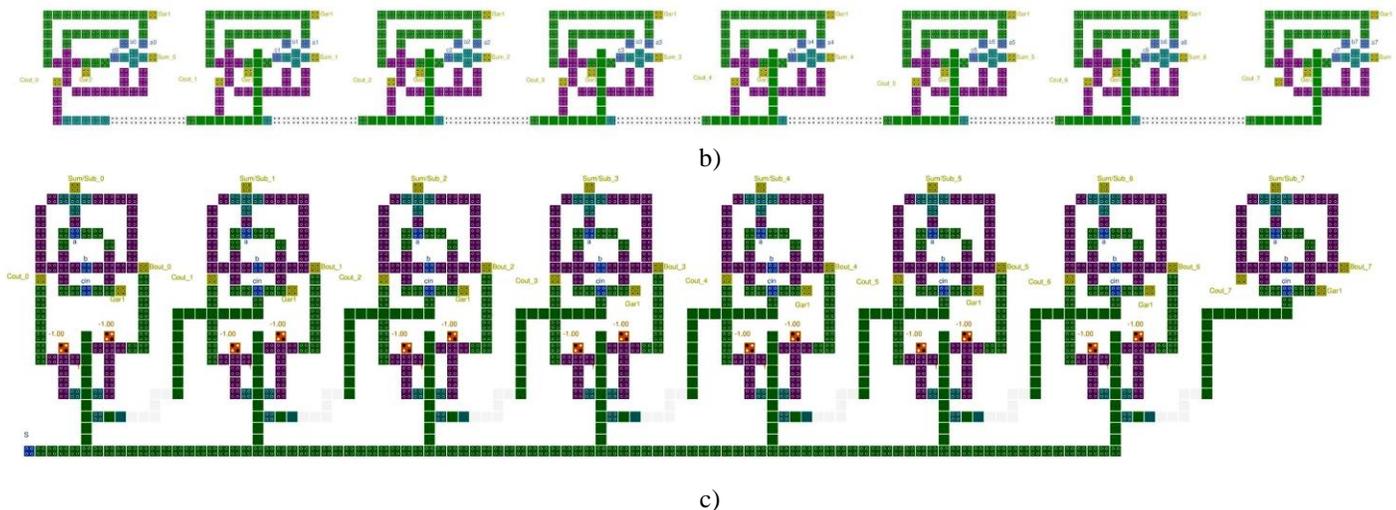

b)

c)

Fig. 4: QCA layout of a) 8-bit reversible full-adder using structure proposed in Fig. 6. b) 8-bit reversible full-adder using structure proposed in Fig. 10. c) 8-bit reversible full-adder/subtractor using structure proposed in Fig. 7.

## IV. RESULTS

In this section, first simulation results of proposed circuits are presented. Then, in the next section, a comparison between the best obtained results are done.

### A. Simulation results

For the proposed circuit layouts and functionality checking, a simulation tool for QCA circuits, QCA Designer version 2.0.3 [39][37] is used in a bistable approximation [4, 18, 40, 41]. QCA Designer is a tool used to create layout and accurate simulation for QCA circuits that can run on most platforms. Parameters used in this software are shown in Table IV.

TABLE IV
SETTINGS OF QCA DESIGNER

| PARAMETER TYPE | VALUE | PARAMETER TYPE | VALUE |
|---|---|---|---|
| CELL SIZE | 18 nm | CLOCK HIGH | 9.8E−22 |
| NUMBER OF SAMPLES | 12800 | CLOCK LOW | 3.8E−23 |
| CONVERGENCE TOLERANCE | 0.001000 | CLOCK AMPLITUDE FACTOR | 2.000 |
| THE RADIUS OF EFFECT | 65 nm | LAYER SEPARATION | 11.5 nm |
| RELATIVE PERMITTIVITY | 12.9 | MAXIMUM ITERATIONS PER SAMPLE | 100 |

Simulation results of proposed reversible designs are shown in Fig. 14. Also, Fig. 15 depicts simulation results of irreversible adder and AND constructed using the five-input majority gate.

### B. Comparison results

The proposed full-adders are compared with the previously proposed ones in reversibility and the results are illustrated in Table V. To demonstrate performance of proposed designs, we compare them with irreversible designs in Table VI. Also, the proposed 8-bit reversible full-adders/subtractors are compared in Table VII. It can be seen that the proposed reversible/irreversible full adders are more efficient in terms of area, complexity, delay, reversible/irreversible layout, and also in logic level in terms of garbage outputs, control inputs, number of majority and NOT gates compared to the best previous works in reversible [29, 30] and irreversible [18, 20-22] and [30, 42] for 8-bit reversible full-adders/subtractors. In fact, Table VI show that our results are even better when results of proposed reversible full-adders/subtractors are compared with section of irreversible adders.

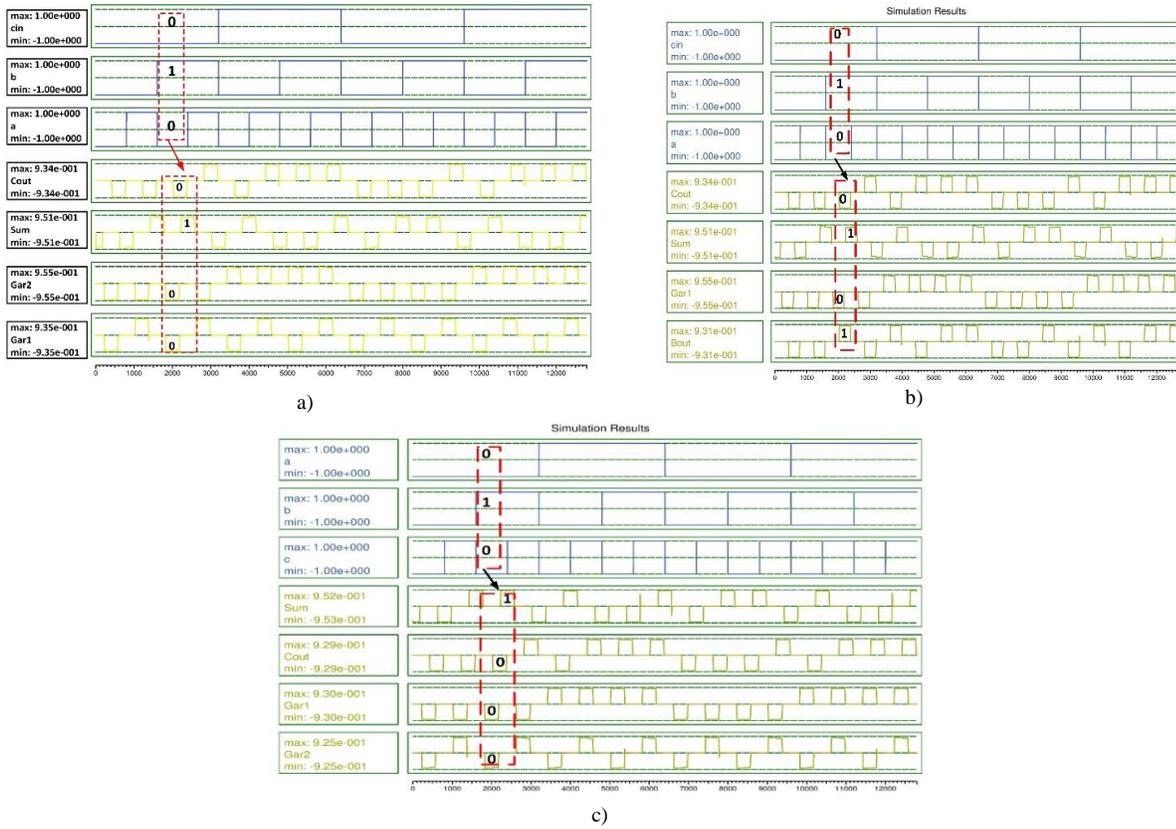

Fig. 5. Simulation results. a) Proposed reversible adder. b) Proposed reversible adder/subtractor. c) Proposed reversible adder with five-input majority

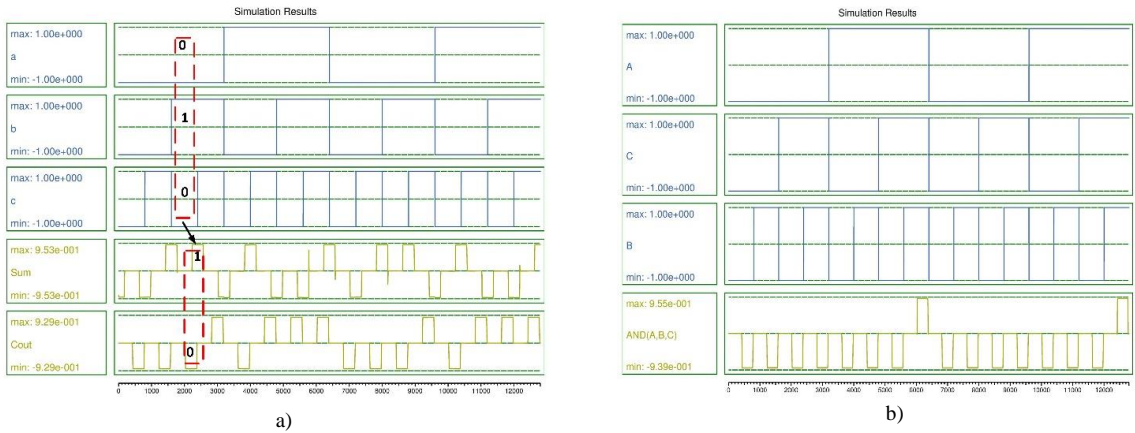

Fig. 6. Simulation results. a) Irreversible Adder proposed with five-input majority gate. b) three-input AND

TABLE V
COMPARISON OF QCA REVERSIBLE FULL-ADDER DESIGNS

| REVERSIBLE FUNCTIONS | CONSTANT INPUTS | GARBAGE OUTPUTS | COMPLEXITY (CELL AMOUNT) | AREA ($\mu m^2$) | CLOCKING ZONES (DELAY) | NUM OF MAJORITY GATES | NUM OF NOT GATES | CONTROL INPUTS |
|---|---|---|---|---|---|---|---|---|
| PREVIOUS DESIGN [30] | 0 | 2 | 399 | 0.5 | 8 | 9 | 13 | 3 |
| PREVIOUS DESIGN [29] | 0 | 3 | >300 | 0.361 | 6 | 6 | 4 | 0 |
| PROPOSED QCA FULL-ADDER | 0 | 2 | 48 | 0.04 | 3 | 3 | 2 | 0 |
| PROPOSED FULL-ADDER/SUB | 0 | 2 | 48 | 0.04 | 3 | 3 | 2 | 0 |
| PROPOSED FULL-ADDER WITH FIVE-INPUT MAJORITY GATE | 0 | 2 | 58 | 0.04 | 3 | 2 | 2 | 0 |

TABLE VI
COMPARISON OF QCA IRREVERSIBLE FULL-ADDER DESIGNS

| IRREVERSIBLE FUNCTIONS | COMPLEXITY (CELL AMOUNT) | AREA ($\mu m^2$) | CLOCKING ZONES (DELAY) | Layers |
|---|---|---|---|---|
| PREVIOUS DESIGN [22] | 145 | 0.17 | 5 | 1 |
| PREVIOUS DESIGN [18] | 86 | 0.10 | 3 | 2 |
| PREVIOUS DESIGN [20] | 73 | 0.04 | 3 | 2 |
| PREVIOUS DESIGN [21] | 52 | 0.04 | 3 | 2 |
| PROPOSED QCA FULL-ADDER | 46 | 0.04 | 3 | 1 |
| PROPOSED FULL-ADDER/SUB | 46 | 0.04 | 3 | 1 |
| PROPOSED FULL-ADDER WITH FIVE-INPUT MAJORITY GATE | 52 | 0.04 | 3 | 1 |

TABLE VII
COMPARISON 8-BIT REVERSIBLE ADDERS/SUBTRACTORS IN QCA

| | COMPLEXITY (CELL AMOUNT) | AREA ($\mu m^2$) | CLOCKING ZONES (DELAY) |
|---|---|---|---|
| PREVIOUS DESIGN (NOT REVERSIBLE) [42] | 5786 | 10.47 | 109 |
| PREVIOUS DESIGN [30] | 5489 | 10.07 | 92 |
| PROPOSED QCA 8-BIT REVERSIBLE FULL ADDER (1) | 570 | 0.55 | 32 |
| PROPOSED QCA 8-BIT REVERSIBLE FULL ADDER (2) | 725 | 0.68 | 32 |
| PROPOSED QCA 8-BIT REVERSIBLE FULL ADDER/SUBTRACTOR | 1040 | 1.12 | 42 |

## V. CONCLUSION

In this paper, Innovation was divided to three section. In the first section, a method for converting irreversible functions to a reversible one was presented. This method had advantages such as: converting of irreversible functions to reversible one directly and as optimal. So, in this method, sub-optimal method of using of conventional reversible blocks such as Toffoli and Fredkin was not used, having of minimum number of garbage outputs and garbage outputs do not create any additional gates (Majority and NOT gates) and so on. Then, Using the method, two new symmetric and planar designs of reversible full-adders were presented. In second section, a new symmetric, planar and fault tolerant five-input majority gate was proposed. Based on the designed gate, a reversible full-adder were presented. Also, for this gate, a fault tolerant analysis was proposed. And in third section, three new 8-bit reversible full-adder/subtractors was designed based on full-adders/subtractors proposed in the second section. The layouts and functionality checks were done using QCA Designer and the designs was compared to the best previous QCA adder designs and shows considerable improvements in terms of area, complexity, delay, reversible/irreversible layout, and also in logic level in terms of garbage outputs, number of majority, NOT gates and power conception due to reversibility. These circuits can be used in processors with high operating speeds.